# Semi-Cooperative Passive Integrated Sensing and Communication by Utilizing Physical Layer Information of 5G Signals

Bo Wei, Ryusei Ogane, and Hang Song

***Abstract*— In recent years, integrated sensing and communication (ISAC) has attracted significant attention towards future cellular networks. Currently, various works have demonstrated sensing performance in existing wireless communication systems. Most of the demonstrations are based on passive type due to the radio regulatory. However, because of the difficulty in access to the communication protocol stacks in commercial cellular systems, the evaluation of the cellular communication signals for semi-cooperative passive ISAC is limited. In this paper, a semi-cooperative passive ISAC system is developed with open-source 5G framework and software-defined radio devices. And the performance is experimentally evaluated by utilizing the physical layer information of actual 5G signals from the developed system. In this system, communication is established with 5G signals and physical layer information are obtained and extracted for analysis. The performance of the system is verified by conducting experiments under several communication scenarios including the synchronization, pinging, and data transmission. Different physical layer information is collected, and the propagation characteristics are analyzed by a multi-path configuration. The experiment results demonstrated that variable bandwidths were utilized for different communication scenarios and the multipath was successfully detected with two different approaches. These results show that the developed system is promising for semi-cooperative passive ISAC in wider application fields.**

## I. Introduction

WITH the spread of the fifth-generation (5G) mobile communication system, researches have been carried out aiming at more effectively utilizing limited radio resources. 5G new radio (NR) employs the orthogonal frequency division multiplexing (OFDM) frame structure and a flexible bandwidth scheme to dynamically adapt the frequency resources and subcarriers spacing for supporting diverse communication services [1]. Along with the advancement of wireless systems, utilizing the communication signals for other purposes has also been attracting wide attention, where sensing is one of the most focused subjects.

Integrated sensing and communication (ISAC) has been widely studied in the past years, which aims at increasing radio resource utilization efficiency [2-4]. In ISAC, the goal is to achieve object detection, ranging and speed estimation, environment identification etc., by utilizing information of the transmitted communication signals. ISAC generally does not require the introduction of new signal generators such as radar or other sensors such as camera, which is promising for low-cost and low-complexity implementation. Besides spectrum efficiency, ISAC also provides an alternative sensing method which does not collect image information as the camera does. Thus, the privacy issue can be reduced. Also, it can also work in low visibility environment such as at night. To realize ISAC, it is necessary not only to evaluate the communication quality in higher layer using the communication signals, but also to analyze them in detail from sensing perspective. Particularly, understanding the channel characteristics in the physical layer is extremely important.

Generally, there are two kinds of ISAC depending on which part is the main structure, communication-centric and sensing-centric [5]. For the sensing-centric approach, the ISAC system is mainly designed for sensing, and also realizes the communication function. While for the communication-centric approach, the signals of existing communication system can be reused for sensing, which does not need to change the current communication structure. By analyzing the signals obtained from communication, the temporal and spectral features can be utilized for characterizing the propagation channel and achieving sensing purpose. Recent years, there are numerous works which focused on different aspects of the ISAC technologies, including the waveform design [6], signal processing algorithms [7, 8], and channel modeling [9, 10].

Although theoretical research possesses an important role in the development of ISAC technologies, there are also a lot of works developing the prototype ISAC system for real-world sensing demonstrations, which show the state-of-the-art of the ISAC implementation [11-14]. According to the cooperation level between the transmitter (Tx) and the receiver (Rx), the ISAC can be categorized into 3 types: non-cooperative passive, semi-cooperative passive, and cooperative active [11]. In non-cooperative passive type, Tx and Rx are basically positioned separately. Rx does not have the information of the transmitted waveform and does not synchronize with the Tx. The Rx only passively receives the signals from the Tx. In semi-cooperative passive type, the Tx and Rx are generally also separately located. But the Rx has the knowledge of the transmitted signals from Tx. Therefore, channel estimation can be conducted. In cooperative active type, the Rx is generally co-located with the Tx and the synchronization is achieved perfectly between the Tx and Rx. In this type, the Tx actively emits the signals for

Bo Wei is with School of Informatics and Data Science, Hiroshima University, Higashi-Hiroshima 739-8527, Japan and also with Japan Science and Technology Agency (JST), PRESTO, Kawaguchi, Saitama 332-0012, Japan.

Ryusei Ogane is with School of Engineering, Okayama University, Okayama 700-8530, Japan.

Hang Song is Research Institute for Semiconductor Engineering, Hiroshima University, Higashi-Hiroshima 739-8527, Japan.

sensing and the Rx has the full information of the transmitted signals.

Currently, many ISAC implementations with existing communication systems are carried out with the passive type considering the restrictions of emitting radio waves. For non-cooperative type, different signals were utilized including WiFi and cellular signals [15-17]. In contrast, most of the semi-cooperative type were implemented with WiFi signals owing to the accessibility of channel state information (CSI) using specific network interface cards (NICs) [18-22]. However, it is difficult to access lower layer information of the cellular network such as signals used in the physical layer and channel estimation results, especially in 5G mobile communication system. Therefore, there are still limitations on flexibly accessing and analyzing the 5G signals at physical layer level for sensing purposes. The semi-cooperative ISAC with actual 5G signals has not been thoroughly investigated yet.

To address the above issue, in this paper, a semi-cooperative passive ISAC system is developed by using the open-source 5G framework [23] and software-defined radio devices. Here, OpenAirInterface (OAI) is chosen because of its compliance with 3GPP standards and its accessibility to the information of different layers during communication, which can be utilized for sensing. Algorithms developed for the semi-cooperative passive ISAC with OAI can be transplanted to existing communication devices. In the developed system, the gNB and UE are implemented separately. Due to the regulation of radio emission, gNB and UE are connected with cables for communication. During communication, all the signaling and data transmission are carried out normally without violating the protocols. Meanwhile, to realize sensing from communication, different kinds of information are extracted including demodulation reference signals (DMRS) of physical broadcast channel (PBCH) and physical download channel (PDSCH) without interrupting the communication. The performance of the semi-cooperative ISAC system is evaluated by analyzing the extracted signals under several communication scenarios including the synchronization, pinging, and large data transmission. In the experiment, the multi-path configuration was set by using coaxial cables with different lengths. The propagation characteristics were analyzed with two different approaches, the power delay profile (PDP) and frequency-domain interference. To the best of our knowledge, this is the first work which systematically analyzes the actual 5G signals towards semi-cooperative ISAC.

The experiment results demonstrated that the variable bandwidths were utilized for different communication scenarios and the multipath was successfully detected. These results show that the developed system is promising for semi-cooperative passive ISAC in wider application fields. The main contributions of this work are summarized as follows.

(1) A semi-cooperative passive ISAC system in 5G is proposed which utilizes the physical layer information from actual 5G signals without interrupting the communication.

(2) Experiments are conducted to evaluate the performance of the proposed system and two approaches are implemented to analyze the results. The comparison is also carried out to show the features of different methods.

(3) The feasibility of the proposed system under different communication conditions is demonstrated by successfully differentiating multiple paths, which is promising for the real-world deployment of semi-cooperative ISAC.

The reminder of this paper is organized as follows. In Section II, the related works are reviewed for the ISAC systems. In Section III, the proposed semi-cooperative ISAC system for 5G is detailed. In Section IV, the signal processing methods are presented. In Section V, the experiment results are discussed to show the performance. Finally, the conclusion is made.

## II. Related Work

Development of the prototype ISAC systems has been widely studied in the past decade. Depending on how the Tx and Rx cooperate, the ISAC system can be generally classified into 3 categories: cooperative active, non-cooperative passive, and semi-cooperative passive.

In the non-cooperative passive ISAC, the Rx only receives Tx signals without demodulation and decoding. This is basically because of the lack of protocol information of the Tx signal. In [11], Liu *et al*. developed a non-cooperative passive ISAC system by using the 25 GHz mmWave communication signals. This mmWave system is for low power data communication with a specific protocol. By utilizing real mmWave communication signals, the indoor human movement speed and pattern have been accurately estimated. Non-cooperative passive ISAC is also carried out with well-known signals such as WiFi. Even though the Tx signals are known, they are treated as fundamental waveforms in the signal processing without extracting the embedded information. Li *et al*. utilized the WiFi signal for human sensing, where the WiFi signals are processed by cross ambiguity function (CAF) to extract Doppler information [16, 17]. Chu *et al*. utilized the digital terrestrial multimedia signals to detect the low-slow-small targets by proposing a deep learning-based network model [24].

In semi-cooperative passive ISAC, the Rx has the information of the transmitted signals. Therefore, more details such as channel state information (CSI) can be extracted through demodulation. WiFi has been widely studied for semi-cooperative passive ISAC. Except for the coarse-grid information such as received signal strength indicator (RSSI), fine-grid CSI of the subcarriers can be extracted by several commercial off-the-shelf (COTS) NICs [18, 19]. By analyzing the CSI dynamics, human sensing has been investigated. Custance *et al*. proposed a transformer network-based gait identification method by using the WiFi CSI [25]. Alzaabi *et al*. utilized the WiFi CSI to monitor the vital signals including the respiratory rate and heart rate of the older people in various environments [26]. Duan *et al*. proposed an intelligent intrusion detection system with WiFi CSI by using a tensor-based approach [27]. Wang *et al*. exploited WiFi CSI from multisensory array for human activity recognition by a deep-learning framework [28]. Liu *et al*. utilized the WiFi CSI as a fingerprint for indoor localization by proposing principal component analysis and graph temporal convolutional network [29]. Besides the dynamics caused by reflection of humans, the transmission characteristics of the WiFi signals have also been utilized for monitoring the change of the propagation environment. Burke *et al*. utilized the WiFi CSI to estimate the humidity via machine learning technique [30]. Li *et al*.

proposed a WiFi sensing system which is deployed in vehicular tunnel to monitor environment safety such as fire accidents [31]. Also, the quantitative measurement of the dielectric properties of liquids is enabled by revealing the relationship between the RSSI and CSI [32, 33].

Although semi-cooperative ISAC has a lot of advantages due to the access of detailed channel information, it is limited to specific communication systems such as WiFi, which can provide such information. In general case, the detailed channel information is not available in the cellular system for users. Only some coarse-grid information such as RSSI can be accessed and the update rate is low in the order of second. Slow-rate coarse information cannot be used for accurate sensing in dynamic scenarios. Therefore, the potential of using cellular signals for semi-cooperative ISAC has not been fully exploited. In contrast, in this paper, the proposed system can extract the physical layer information of actual 5G signals and the extracted information is analyzed for the purpose of semi-cooperative ISAC.

## III. 5G Semi-Cooperative Passive ISAC

In this paper, the physical layer information of the actual 5G signals is utilized for semi-cooperative ISAC. 5G NR (new radio) is now widely deployed which enables high-speed, low-latency, and massive-connected communications. The physical layer of 5G is based on orthogonal frequency division multiplexing (OFDM) and the resources are flexibly assigned in both temporal and frequency domain. Depending on the communication environment and service requirement, the physical layer is dynamically adjusted. Theoretically, by analyzing the physical layer signals through demodulation, the propagation channel status can be estimated. Then the sensing function can be enabled from the channel status.

### *A. Reference Signals in 5G*

In 5G, the channel is estimated through reference signals (RS). There are different kinds of RS in both downlink and uplink, such as demodulation reference signal (DM-RS), channel state information reference signal (CSI-RS). This work mainly focuses on the DM-RS in downlink to develop the semi-cooperative ISAC, by which no extra spectrum resource is required and only the existing communication radio resource is reused. DM-RS is utilized in physical downlink shared channel (PDSCH) associated with data transmission. DM-RS is also utilized in physical broadcasting channel (PBCH) associated with the synchronization signal block (SSB).

The SSB is emitted from the gNB periodically for synchronization with UE. SSB occupies 20 resource blocks (RB) which is a fixed structure. And 12 subcarriers form one RB. Therefore, there are 240 subcarrier frequencies that can be used. SSB consists of 4 OFDM symbols. In symbol 0, primary synchronization signal (PSS) is embedded which occupies 127 subcarriers. Symbols 1 and 3 are all for PBCH. In symbol 2, the second synchronization signal (SSS) and some parts of PBCH share the resource elements (REs). During the synchronization, the UE first extracts the cell ID value. Consequently, the position of the DM-RS can be derived from the cell ID value. Then, the channel estimation is carried out by using the DM-RS and the PBCH is demodulated. In contrast, the occupied RB in PDSCH is dynamically changed depending on the services. The DM-RS is embedded in PDSCH together with the data blocks. The position where DM-RS is allocated is controlled by several parameters including mapping type and configuration type. By using the channel estimation from DM-RS of PDSCH, the data conveyed in PDSCH can be recovered.

Depending on the associated physical channel, DM-RS shows different features. In PBCH, DM-RS is fixed in both frequency and time domain. Since SSB is periodic, the DM-RS in PBCH is also periodic. While in PDSCH, DM-RS is dynamic in both frequency and time domain. Since PDSCH changes depend on conditions and services, the size of DM-RS also varies accordingly. The change of size in frequency domain directly influences the bandwidth, as well as the sensing ability. In this work, both the PBCH and PDSCH are investigated for comparison.

### *B. Channel Estimation by DM-RS*

The channel can be estimated by comparing the received signal and the emitted signal. Denote the transmitted DM-RS from gNB as $X_{\mathrm{DMRS}}[k]$ in frequency domain. $k$ is the index of subcarrier frequency. Then, the channel state information is calculated as

$$H_{\mathrm{DMRS}}[k] = \frac{Y_{\mathrm{DMRS}}[k]}{X_{\mathrm{DMRS}}[k]} \tag{1}$$

where $Y_{\mathrm{DMRS}}[k]$ is the received signal in frequency domain after Fourier transform. Since the pattern of DM-RS is already determined in the 5G protocol which is $X_{\mathrm{DMRS}}[k]$, the channel status at each subcarrier $H_{\mathrm{DMRS}}[k]$ is obtained. $H_{\mathrm{DMRS}}[k]$ can not only be utilized for demodulating the data, it can be also utilized for sensing by investigating the channel status. During communication, when the radio wave propagates in multipaths, or if there is any movement introduced by human or other objectives, the channel will change accordingly and it will be reflected in the channel state information.

### *C. Semi-Cooperative ISAC system*

The semi-cooperative ISAC system was developed by incooporating the OAI framework. OAI is a software-based implemenetation of 5G NR protocol stacks defined in 3GPP standards. With the OAI, the gNB, UE and core networks (CN) functions can be deployed locally with computers and software-defined radio (SDR). The reason for choosing OAI is that it provides full access to the communcation stacks. Therefore, it becomes possible to extract all kinds of physical layer signals and the channel estimation results, which are typically not reachable in commercial 5G products. This enables the analysis of phyiscal layer signals and development of new algorithms and modules to extend the communication signals for sensing.

The system structure is shown in Fig. 1. The system is configured as one gNB and one UE. Two SDRs (USRP B210, Ettus, USA) are utilized as the radio wave transceiver at both the gNB and UE sides. On the gNB side, both the gNB and CN are implemented on one general-purpose computer. The UE is implemented on another computer. OAI stacks are deployed separately on gNB and UE sides. In the developed system, the ISAC is performed on UE side by using the downlink signals.

To experimentally evaluate the semi-cooperative ISAC, both the uplink and downlink of the communication are established by connecting with coaxial cables. Since the downlink is mainly considered in this work, the uplink is directly connected with one cable. The attenuator is also inserted to adjust the radio wave power. To evaluate the performance of the ISAC in downlink, a multipath environment is configured to split the transmitted signal from the gNB into multiple paths by using the power divider and power combiner. One path is a short one with length of 0.6 m. Other paths are longer ones with the length of several tens meters. Another atteuator is also inserted in the short path to adjust the received power balance among paths.

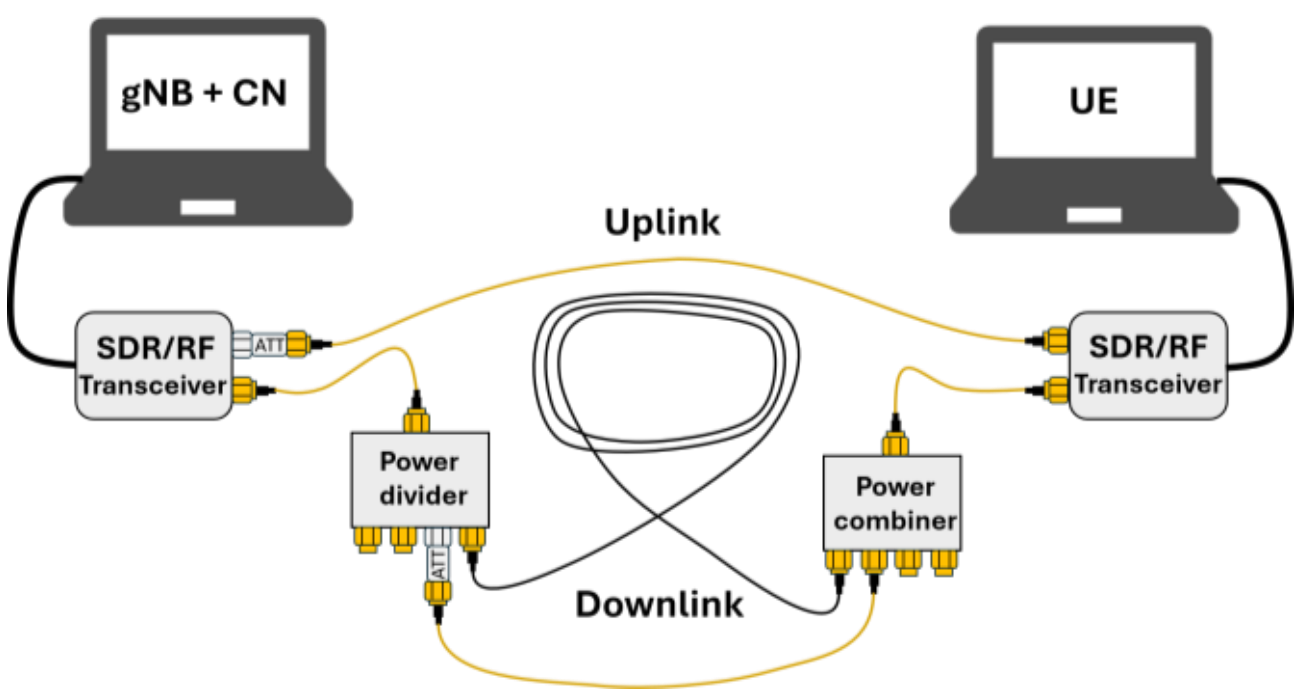

**Fig. 1** Semi-cooperative ISAC system structure.

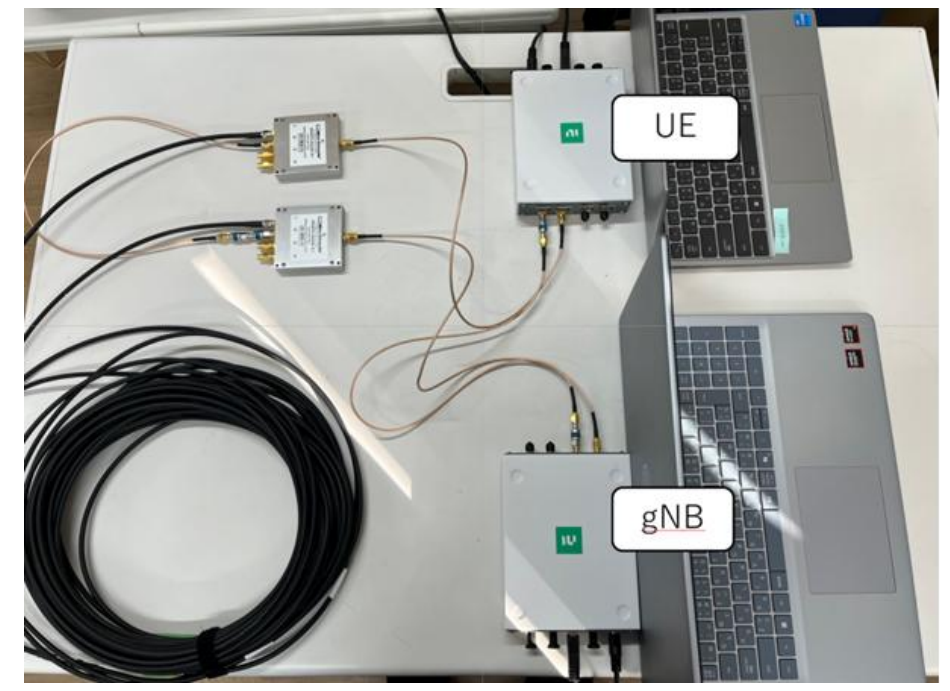

**Fig. 2** Semi-cooperative ISAC system under operation.

With the proposed system, the semi-cooperative passive ISAC can be experimentally verified by utilizing physical layer information of actual 5G signals as shown in Fig. 2. In the experiment, the communication are established under different scenarios including synchronization, pinging, and high-rate data transmission. Then, by introducing the channel data extraction and analysis modules on UE side, the DMRS data are retrieved and the multipath effect are analyzed.

## IV. Signal Analysis Methods

To analyze the received signals in multipath configuration, two methods are utilized on the UE side. One is the power delay profile (PDP) in time domain. The other is superposition of waves in frequency domain.

### A. Power Delay Profile

PDP is utilized to investigate the multipath effect during communication. It reflects the difference of arrival delays from various propagation paths. PDP can be calculated from the frequency response. Denote the channel response estimated on UE side as $H[k]$. Then, the impulse response can be calculated by:

$$h[n] = IFFT\{H[k]\} = \frac{1}{N}\sum_{k=0}^{N-1} H[k]e^{j2\pi\left(\frac{kn}{N}\right)} \quad (2)$$

where $n$ is the discrete time index. $\boldsymbol{N}$ is the number of the subcarriers. Denote the subcarrier spacing as $\Delta f_{\mathrm{SCS}}$. Then, the bandwidth is calculated as $B = N\Delta f_{SCS}$ and the time resolution $\Delta t$ is determined by the reciprocal of $B$. To suppress the sidelobe of Sinc function, a window function $W(k)$ can be applied to $H[k]$ before the inverse Fourier transform as:

$$H_{\mathrm{w}}[k] = H(k) \cdot W(k) \quad (3)$$

Then, $H[k]$ in Eq. (2) can be replaced by $H_{\mathrm{w}}[k]$. On the other hand, the time resolution will deteriorate depending on the window function used. Finally, the PDP is calculated by taking the square of $h[n]$ as:

$$PDP[n] = |h[n]|^2 \quad (4)$$

By examining the peaks in PDP, the arrival time of multiple paths can be derived. In this work, the utilized bandwidth varies in different communication scenarios. Therefore, the time resolution also varies in different scenarios.

### B. Superposition of Waves

In the frequency domain, the multipath components can also be evaluated by the superposition of waves from different paths. Especially, in the two-path configuration, the signals arrive at the receiver sides with different delays. Since the structure of the signals are the same, the inference among signals will occur at each subcarrier frequency. Considering a channel with two paths, the frequency-domain channel response at $k$th subcarrier can be expressed as:

$$H(k) = a_1 + a_2 e^{-j2\pi k\Delta f_{\mathrm{SCS}}\Delta\tau} \quad (5)$$

where $a_1, a_2$ are the complex amplitudes of each path, and $\Delta\tau$ is the delay difference between the two paths. $e^{-j2\pi k\Delta f_{\mathrm{SCS}}\Delta\tau}$ affects the phase of second path and the second term will rotate in the complex plane. Since the subcarrier frequency changes with the same spacing, $|H(k)|^2$ exhibits periodic fluctuations along the frequency axis according due to the constructive and destructive interference in the correlation. The relationship between the period $\Delta f$ of the fluctuation in $|H(k)|^2$ caused by interference and $\Delta\tau$ is approximately given by:

$$\Delta f \approx \frac{1}{\Delta\tau} \quad (6)$$

By examining the period of the fluctuation in the channel response, it is possible to derive the time difference between two paths.

## V. Experimental Results

To evaluate the performance of the proposed semi-cooperative passive ISAC system, several experiments were conducted. In the experiments, the USRP B210 SDR was utilized which can support maximum 56 MHz bandwidth with the sampling rate of 61.44 MS/s. The subcarrier spacing is set as 30 kHz. To realize relatively stable communication, the maximum RB is set as 106, which is a valid value defined in 5G standard. This is because the SDR and computer were connected by a USB3.0 cable which can transmit the data with rate of 5.0 Gbps. However, this is a theoretical ideal value. To make the data transmission stable, the rate should be set smaller than it. Therefore, the sampling rate is reduced to 3/4, which is 46.08 MS/s, which covers 1536 subcarrier frequencies. Since one RB consists of 12 subcarriers, the maximum effective subcarrier number is 1272 and the bandwidth equals 38.16 MHz. 1536 is larger than 1272, which means that all the effective subcarriers can be included and communication will not be affected by the down sampling.

### *A. Bandwidth and PDP*

In the experiments, three communication scenarios were tested. In the synchronization stage, the SSB signal was captured and the channel response estimated from DMRS of PBCH was extracted. Since the PBCH occupies the full resource blocks of symbol 1 and 3 in the SSB signal, the bandwidth and PDP can be observed by investigating the channel state information in one symbol. Figure 3(a) shows the amplitude of the channel response from PBCH. It can be observed that there are 240 effective subcarriers and the corresponding bandwidth is 7.2 MHz. By taking the inverse Fourier transform of the channel response, the PDP can be obtained as shown in Fig. 4(a). Although there are only 240 subcarriers effective in PBCH, 8192 points were utilized in the transform by using the zero padding. It can be observed that the width of the main lobe is relatively broad due to the narrow bandwidth of the SSB signal.

After the synchronization, the ping requests were sent regularly from UE and the channel response estimated from DMRS of PDSCH was extracted. Figure 3(b) shows the amplitude of the channel response from PDSCH during ping. It can be observed that there are 576 effective subcarriers and the corresponding bandwidth is 17.28 MHz. Calculating the PDP with the same 8192 points by zero padding, the PDP result is shown in Fig. 4(b). It can be observed that the main lobe is narrower compared with that of PBCH case, owing to wider bandwidth.

Since the time resolution in PDP is inversely proportional to the signal bandwidth, the high-rate data transmission was conducted to fully utilize the communication bandwidth and obtain a finer time resolution. In this case, iPerf3 was utilized. The data was transmitted to UE via downlink. During the transmission, the channel response estimated from DMRS of PDSCH was extracted. As shown in Fig. 3(c), there are 1272 effective subcarriers in the channel response which is corresponding to the maximum 106 RB. The PDP result is shown in Fig. 4(b). Since the maximum possible bandwidth was utilized, the time resolution is much better than others.

### *B. PDP in two-path configuration*

The performance of the semi-cooperative ISAC system was also evaluated in the two-path configuration. To get a fine time resolution, the communication scenario was chosen to be the high-rate data transmission where the bandwidth was fully utilized. As shown in Fig. 2, one path is connected by a 0.6 meter cable and another is connected with longer cables in this configuration. In the experiment, three 10 meter cables were utilized. The received power was first measured by only connecting with either a 0.6 m cable or a 30 m cable. It was found there was 20 dB difference in power attenuation between the two paths. And each 10 m cable has the insertion loss of around 7 dB.

The experiment was first conducted by using the two-path configuration with the 0.6 m and 30 m cables. In the experiment, a 20 dB attenuator was inserted in the 0.6 m cable path intentionally. The purpose was to make a balanced condition for evaluating the performance. The iPerf3 was utilized to conduct high-rate data transmission to UE. On the UE side, the channel response estimated from DMRS of PDSCH was extracted and the PDP was calculated. Figure 5(a) shows the PDP. It can be observed that there are two peaks in the delay profile and the power is almost the same. This is consistent with the balanced setup of the two paths. The delay difference between the two peaks is around 0.114 μs. Since the path difference is 29.4 m and the nominal velocity of propagation within the cable is 83% of the speed of light, the theoretical time difference should be 0.118 μs. The measured time agree well with theoretical value. Figure 5(b) shows the channel response. It can be observed that the amplitude of the channel response repeated along with the subcarrier frequency. This is consistent with the analysis of wave superposition of the two paths. The average repetition rate is around 283 subcarriers which corresponds to 8.49 MHz, thus the delay difference is estimated as 0.118 μs by Eq. (6). This also agrees well with the theoretical value. Since the two paths are balanced, the amplitude of channel response becomes low at some subcarrier frequency due to the destructive addition of the waves from two paths.

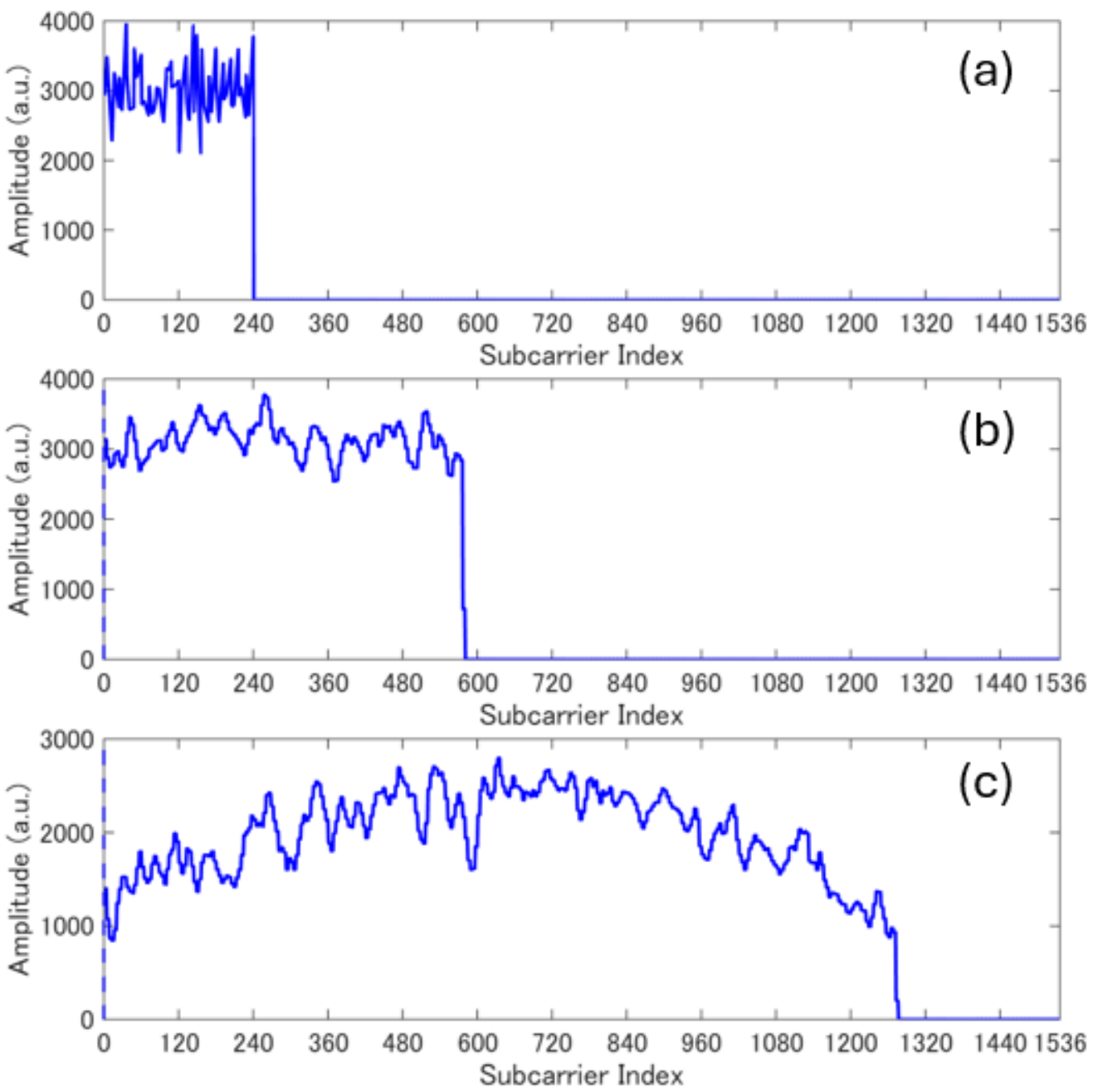

**Fig. 3** Effective subcarriers in channel response estimated from DMRS of (a) PBCH, (b) PDSCH during ping, and (c) PDSCH during high-rate data transmission.

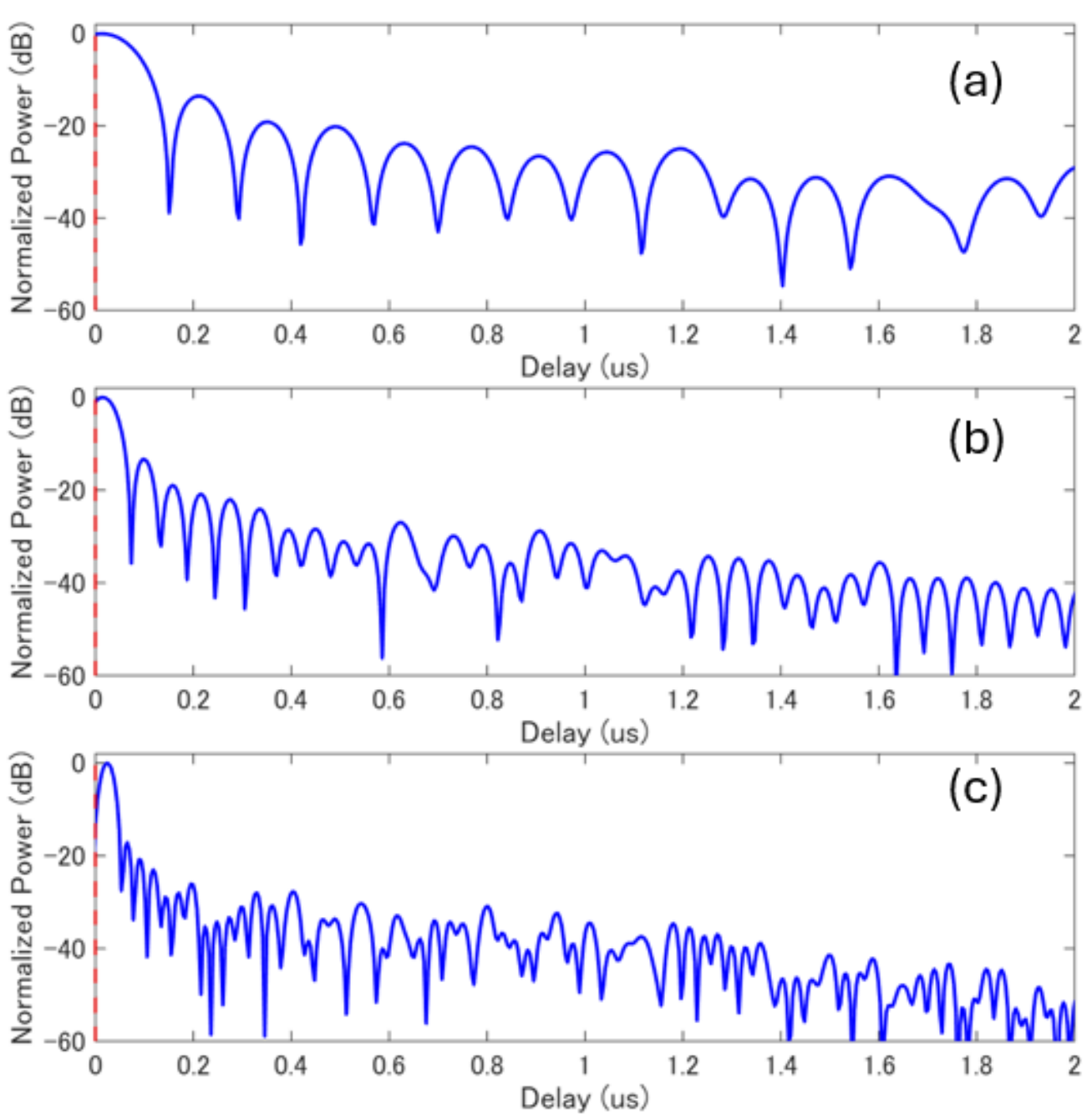

**Fig. 4** PDP results calculated by the channel response estimated from DMRS of (a) PBCH, (b) PDSCH during ping, and (c) PDSCH during high-rate data transmission.

The cable was changed to 20 m in the longer path and the attenuator was changed to 9 dB in the shorter path. The PDP and estimated channel response are shown in Figure 6(a) and (b), respectively. It can be observed that the two peaks are still separatable from the PDP and the delay difference is about 0.073 μs. Considering the path difference is 19.4 m, the theoretical delay difference is 0.078 μs. By investigating the

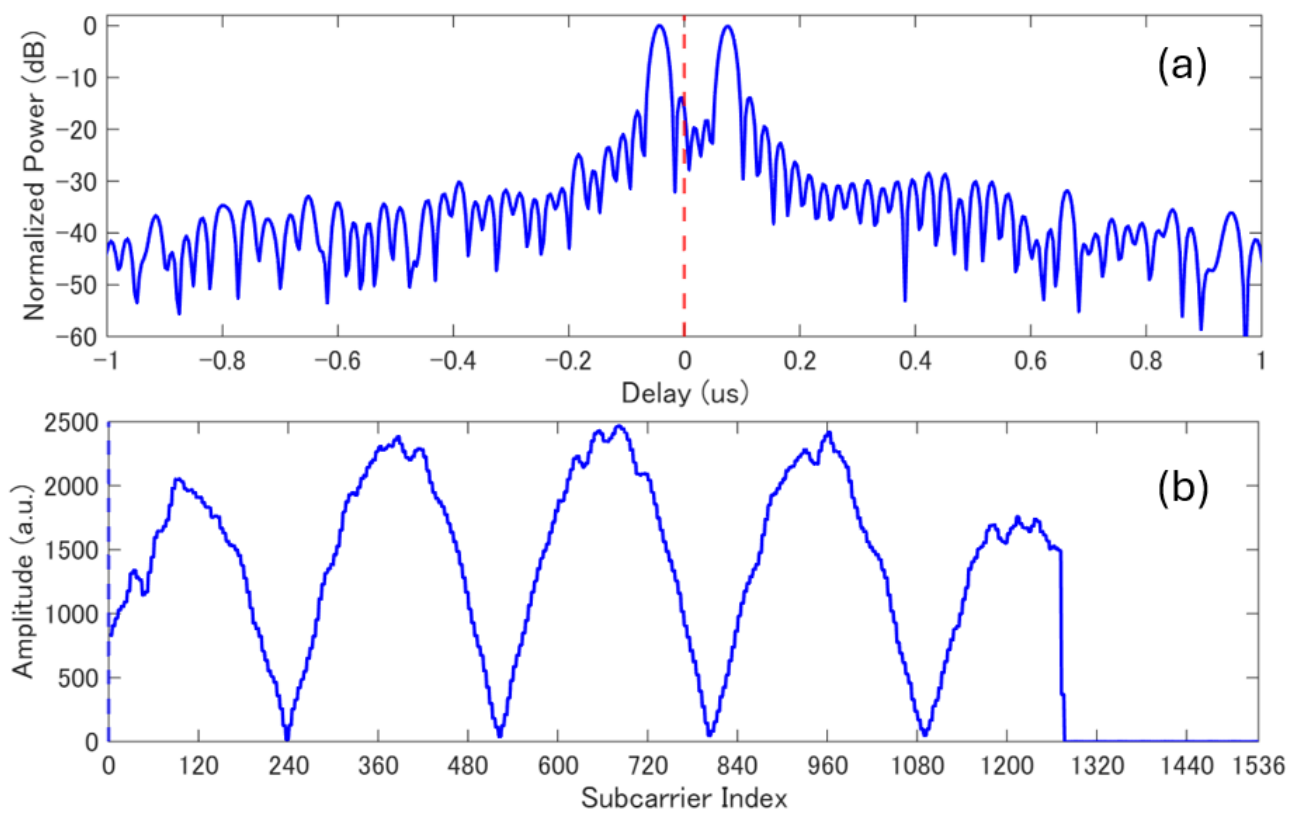

**Fig. 5** (a) PDP result and (b) channel response from DMRS of PDSCH in two-path configuration with 0.6 m and 30 m cables.

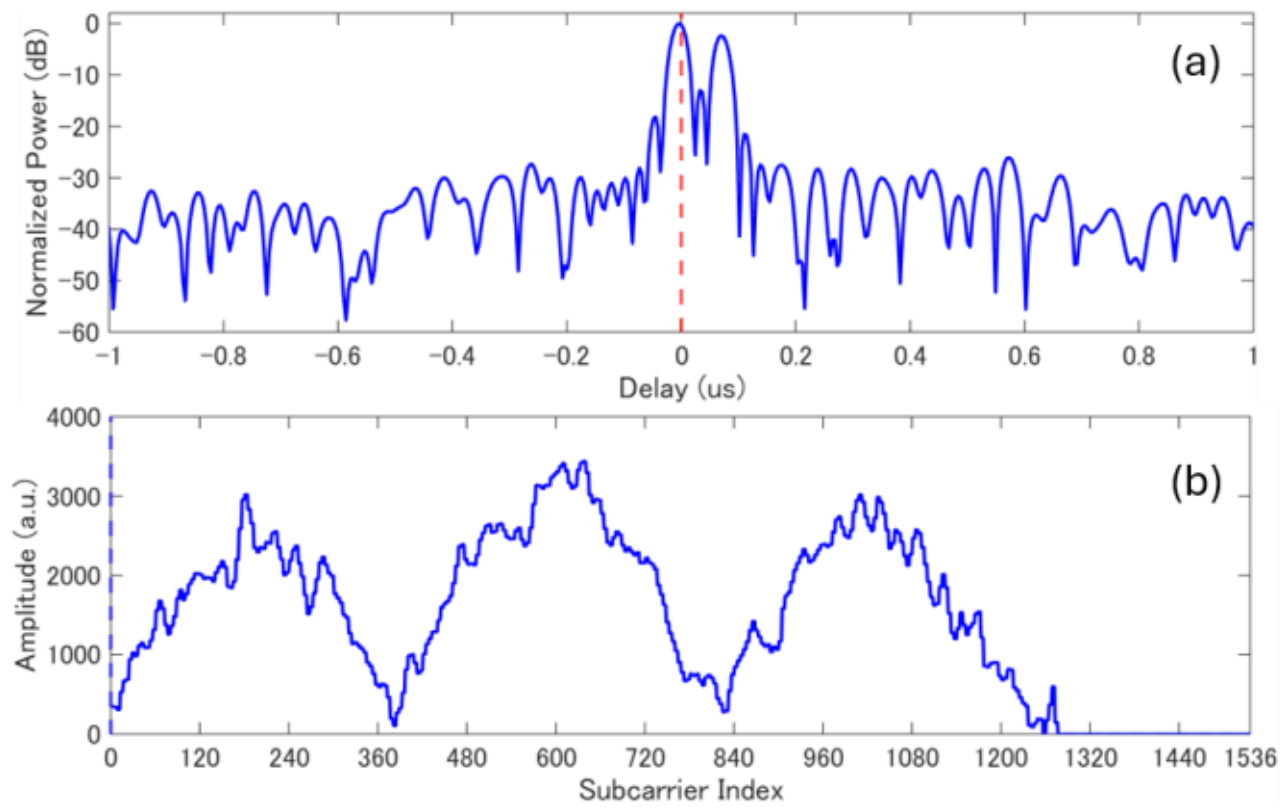

**Fig. 6** (a) PDP result and (b) channel response from DMRS of PDSCH in two-path configuration with 0.6 m and 20 m cables.

channel response, the amplitude also repeated and the repetition rate is around 438 subcarriers which corresponds to 13.14 MHz. Then, the delay difference is estimated to be 0.076 μs. The estimated delay differences from PDP and the channel response both agree well with the theoretical value.

Then, the cable was changed to 10 m in the longer path and the attenuator was changed to 16 dB in the shorter path. This configuration did not consider the balance between two paths. From the PDP results in Fig.7(a), it can be observed that the second peak is more obvious. Since the path difference reduced, the delay difference is also smaller. Although it was still possible to separate the two peaks, the first peak is weak and may be missed if the path difference information was not given in advance. The delay difference between two peaks is 0.037 μs from PDP. Since the path difference is 9.4 m which corresponds to the delay of 0.038 μs, both the value from PDP and theory matched well. On the other hand, from the channel response in Fig.7(b), it was difficult to recognize the repetition of amplitude.

From the results in the two-path configuration, it is found that the detection of different paths is possible by using the DMRS of PDSCH in the proposed semi-cooperative passive ISAC system, which can enable the sensing of different objects in field deployment.

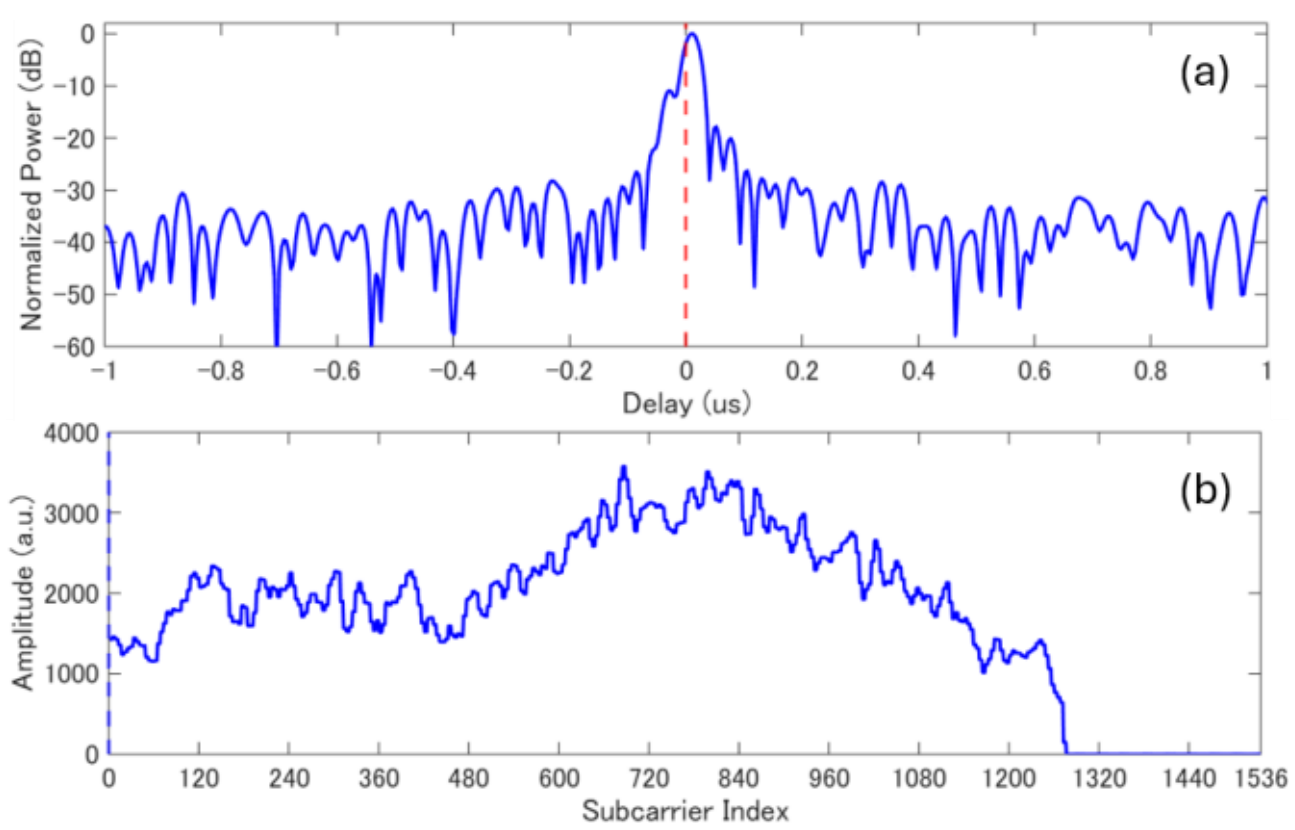

**Fig. 7** (a) PDP result and (b) channel response from DMRS of PDSCH in two-path configuration with 0.6 m and 10 m cables.

## C. *PDP in multi-path configuration*

The multi-path performance was evaluated by setting up three paths with 10 m, 20 m, and 0.6 m cables. An attenuator of 13 dB is inserted in the 0.6 m path. The experiment was conducted under the same conditions as the former ones, where the high-rate data transmission was executed. The PDP result is shown in Fig. 8. There are three peaks observed in the PDP. The delay difference between the first and second peaks is 0.037 μs which agree with the theoretical value. The delay difference between the second and third peaks is 0.045 μs while the theoretical value is 0.040 μs. The error is considered to be caused by the time resolution of the PDP. Since 8192 points were utilized for inverse Fourier transform, the time resolution $\Delta t$ is around 4.1 ns. Therefore, if the peak is chosen with one $\Delta t$ offset, an error of 4.1 ns occurs. The result in multi-path case demonstrated that it is possible to detect several peaks by utilizing physical layer information of 5G signals in the proposed semi-cooperative passive ISAC system.

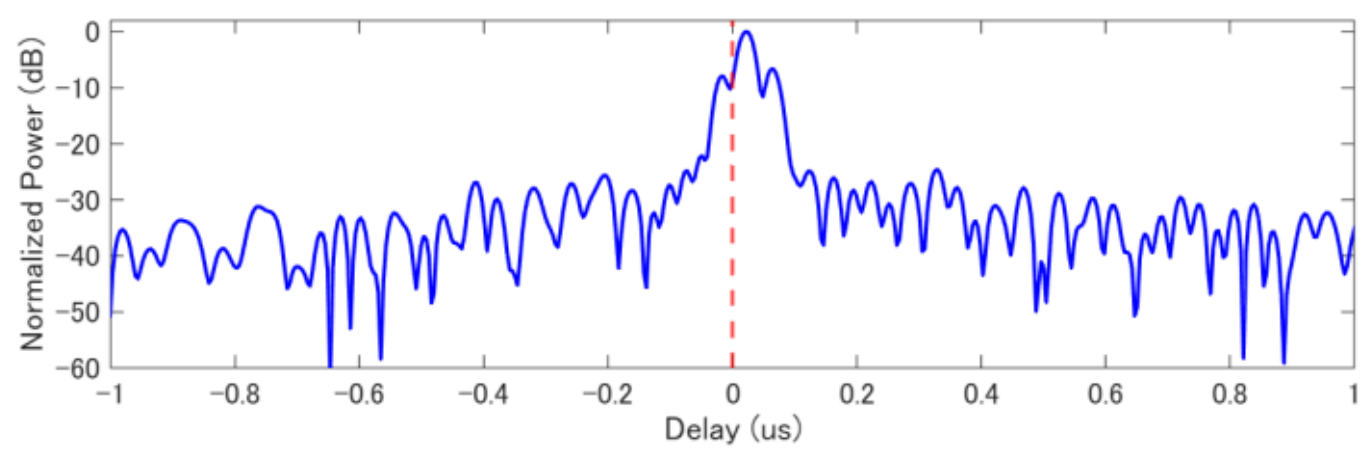

**Fig. 8** PDP result from DMRS of PDSCH in multi-path configuration with 0.6 m, 10 m, and 20 m cables.

## VI. Conclusion

This paper proposed a semi-cooperative passive ISAC system by utilizing physical layer information of 5G signals. In the proposed system, the DMRS of the downlink is accessed by modifying the open-source OAI framework. Then, the channel response is extracted and analyzed in frequency domain. The PDP is also calculated to evaluate the possibility for sensing. Since no extra frames or signals are introduced in the developed system, it is totally passive and compliant with the communication protocols. The performance was evaluated by communication experiments in different scenarios and multi-path configurations. From the results, it was found that the utilized bandwidth is various for different communication scenarios. It is possible to differentiate multiple paths when the high-rate data transmission is carried out. When the bandwidth is 38.16 MHz, the path difference of 10 m is possible to be detected via the experiment. The proposed semi-cooperative passive ISAC system can realize the sensing purpose without interrupting the communication and can be deployed with the existing communication system, which is promising to enable ISAC in wider field applications.